\documentclass[letter]{aa}  
\usepackage{graphicx}
\usepackage{multirow}
\usepackage{txfonts}
\usepackage[]{hyperref}
\begin{document} 
   \title{Deep silence: radio properties of little red dots}
   \author{K. Perger\inst{1,2}         
          \and
          J. Fogasy\inst{1,2}
           \and
          S. Frey\inst{1,2,3}
          \and
          K. \'{E}. Gab\'{a}nyi\inst{4,5,1,2}
          }
   \institute{
             Konkoly Observatory, HUN-REN Research Centre for Astronomy and Earth Sciences, Konkoly-Thege Mikl\'os \'ut 15-17, 1121 Budapest, Hungary  
              \email{perger.krisztina@csfk.org}             
    \and    CSFK, MTA Centre of Excellence, Konkoly-Thege Mikl\'os \'ut 15-17, 1121 Budapest, Hungary   
    \and    Institute of Physics and Astronomy, ELTE E\"otv\"os Lor\'and University, P\'azm\'any P\'eter s\'et\'any 1/A, 1117 Budapest, Hungary
        \and Department of  Astronomy, Institute of Physics and Astronomy, ELTE E\"otv\"os Lor\'and University, P\'azm\'any P\'eter s\'et\'any 1/A, 1117 Budapest, Hungary
        \and   HUN-REN--ELTE Extragalactic Astrophysics Research Group, ELTE E\"otv\"os Lor\'and University, P\'azm\'any P\'eter s\'et\'any 1/A, 1117 Budapest, Hungary      
          }
   \date{Received September 30; accepted xx}
 
  \abstract{To investigate the radio properties of the recently found high-redshift population, we collected a sample of $919$  little red dots (LRDs) from the literature. By cross-matching their coordinates with the radio catalogues based on the first- and second-epoch observations of the Very Large Array Sky Survey (VLASS) and the Faint Images of the Radio Sky at Twenty-centimeters (FIRST) survey, we found no radio counterparts coinciding with any of the LRDs. To uncover possible sub-mJy level weak radio emission, we performed mean and median image stacking analyses of empty-field `Quick Look’ VLASS and FIRST image cutouts centred on the LRD positions.  We found no radio emission above  $3\sigma$ noise levels ($\sim11$ and  $\sim18~\mu$Jy~beam$^{-1}$ for the VLASS and FIRST maps, respectively) in either of the stacked images for the LRD sample,  while the noise levels of the single-epoch images are comparable to those found earlier in the stacking of high-redshift radio-quiet active galactic nuclei (AGNs). The non-detection of radio emission in LRDs suggests these sources host weaker (or no) radio AGNs.}


   \maketitle


\section{Introduction}
\label{Introduction}
One of the surprising results of recent \textit{James Webb Space Telescope} (\textit{JWST}) observations is the discovery of a new population of high-redshift ($z\ge4$) sources, referred to as little red dots \citep[LRDs, e.g.][]{2023ApJ...952..142F,2023ApJ...959...39H,2023arXiv230607320L,2023Natur.616..266L,2024ApJ...963..128B,2024ApJ...964...39G,2024arXiv240403576K,2024ApJ...963..129M}. The common characteristics of LRDs are their compactness and very red continua at rest-frame optical wavelengths, while also exhibiting a blue ultraviolet (UV) excess. Some LRDs also have broad H$\alpha$ emission lines identified in their spectra, indicating that they harbour type-I active galactic nuclei \citep[AGNs, e.g.][]{2023ApJ...959...39H,2024ApJ...964...39G,2024ApJ...963..129M}.

Regarding their nature, there are several theories which try to explain their observed properties with varying success. The red continuum of LRDs could arise from compact, dusty star formation or even from older stellar populations with the stellar UV emission escaping unattenuated from the host galaxy due to uneven dust distribution \citep[e.g.][]{2024arXiv240807745B,2024ApJ...968....4P,2024ApJ...968...34W}. Alternatively, the hot dust emission and the rest-frame optical slope can be explained by a heavily reddened broad-line AGN, with the UV emission originating from either the scattering of the unabsorbed AGN continuum or from their host galaxies \citep[e.g.][]{2024arXiv240610329D,2024ApJ...964...39G,2024arXiv240403576K}. However, the truth probably lies between these two scenarios, as the physical mechanisms that can trigger AGNs can also trigger starburst events, thus it is somewhat expected to see both the central supermassive black hole and the surrounding galaxy grow alongside each other \citep{2024arXiv240610341A}.

Understanding the nature of LRDs is further complicated by the different selection techniques used to identify these sources \citep{2024ApJ...968....4P}, and also by the possible diversity of the newly discovered distant and red galaxies. As \citet{2024arXiv240704777K} pointed out, LRDs are not necessarily H$\alpha$ emitters and vice versa, implying that the LRDs with and without H$\alpha$ emission could represent different galaxy populations. 

A puzzling aspect of LRDs is the lack of X-ray emission, even for sources selected to show broad H$\alpha$ emission lines \citep{2024arXiv240500504M, 2024ApJ...974L..26Y}. To date, only two LRDs have X-ray counterparts, and even the stacking of the \textit{Chandra} fields of X-ray-undetected LRDs did not detect X-ray emission of LRDs \citep{2024ApJ...969L..18A,2024arXiv240403576K, 2024ApJ...974L..26Y}. The lack of X-ray detection could be explained by the intrinsic X-ray weakness of LRDs or heavy X-ray absorption by large column density clouds \citep{2024arXiv240500504M}.
In addition, LRDs selected from the COSMOS-Web  survey \citep{2023ApJ...954...31C} do not have detections at mid- and far-infrared and submillimetre/mm wavelengths either, not even after stacking \citep{2024arXiv240610341A}. If at least a fraction of LRDs hosted AGNs, this would be expected to be reflected in their multi-wavelength properties. Observations in the radio regime are of special value because the propagation of radio waves is not affected by dust and gas obscuration.

To further investigate the nature of LRDs in the scope of AGN activity, we collected an extended list of LRDs from the literature, to explore their radio properties in large radio sky surveys by means of image stacking. In Sect.~\ref{Selection}, we introduce our sample and the cross-match with radio catalogues. In Sect.~\ref{Results}, we describe the radio image stacking analysis and present its results.  Finally, we summarise our findings and conclude the paper in Sect.~\ref{summary}.

\section{Sample selection and radio cross-match}
\label{Selection}

A total of $919$ LRDs were collected from the literature at declinations $\delta>-40\degr$ \citep{2023A&A...677A.145U,2023ApJ...957L...7K,2023ApJ...959...39H,2023arXiv230607320L,2023arXiv230801230M,2023arXiv231203065K,2023Natur.616..266L,2024ApJ...963..128B,2024ApJ...963..129M,2024ApJ...964...39G,2024ApJ...968....4P,2024ApJ...968...34W,2024ApJ...968...38K,2024ApJ...969L..13W,2024arXiv240403576K,2024arXiv240610341A,2024Natur.628...57F}. The list of the selected sources is provided as a machine-readable supplementary material in comma-separated values (\textsc{csv}) format, containing the source identifier (ID) from the published papers where they were presented, the right ascension and declination coordinates in decimal degrees, their spectroscopic and/or photometric redshifts when available, accompanied by the references to main discovery and analysis papers using their SAO/NASA Astrophysics Data System\footnote{\url{https://ui.adsabs.harvard.edu/}} (ADS) bibliographic identifiers. Additional information is provided in the `notes' column when either of the redshift values was collected from an individual publication. 

\begin{figure}[ht]
   \centering
   \includegraphics[width=\linewidth]{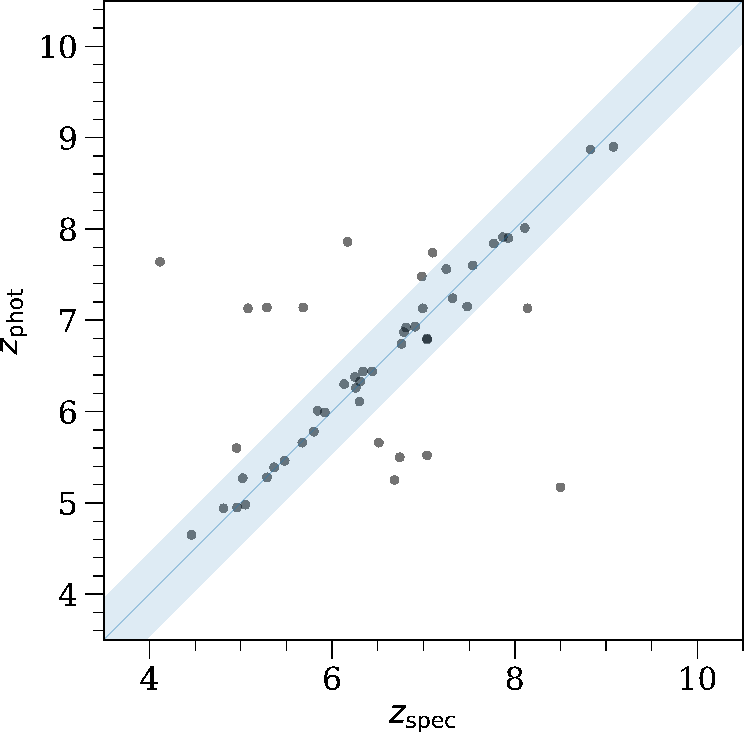}
   \caption{Relation of the spectroscopic and photometric redshift for the $51$ LRDs with both values available. The blue shaded area denotes parity with $\pm3\sigma$ standard deviation.} \label{fig:redshifts}%
    \end{figure}

\begin{figure}[ht]
   \centering
   \includegraphics[width=\linewidth]{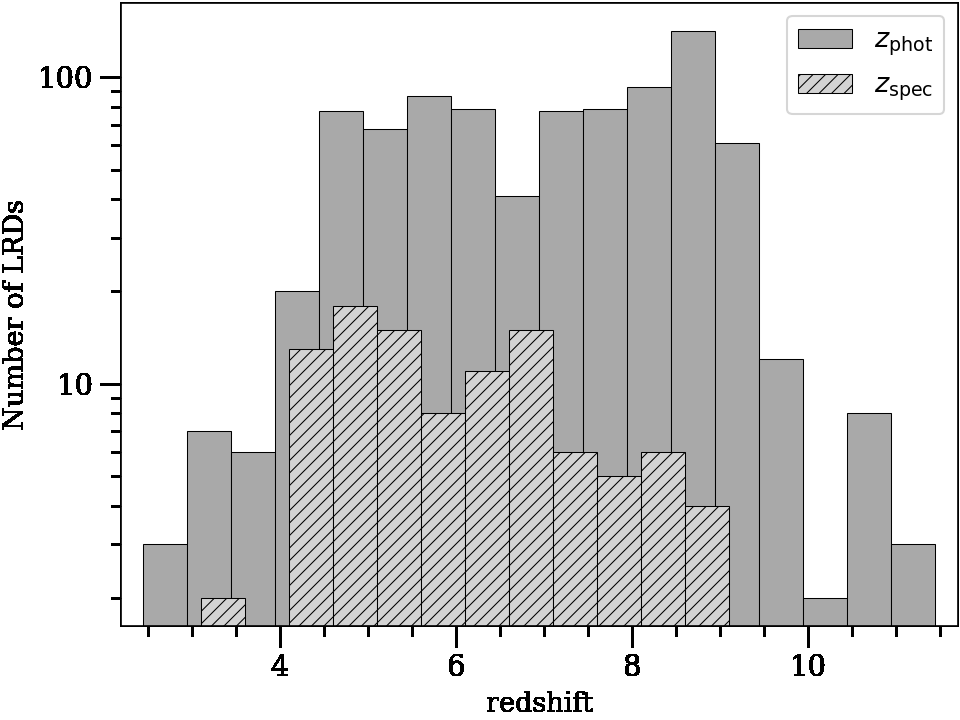}
   \caption{Distribution of LRDs as a function of redshift. Photometric and spectroscopic values are denoted with dark grey and light grey (hatched) coloured bars.} \label{fig:redshift_dist}%
    \end{figure}

The relationship between the spectroscopic and the photometric redshifts is shown in Fig.~\ref{fig:redshifts} for the $51$ LRDs for which both values were available in the literature. Although the similarities are already noticeable visually with most of the objects following the equality line with the slope of unity, statistical parameters of the population also imply the goodness of the photometric redshift estimates. The majority of the LRDs in the sample ($867$ objects) presented here has photometric redshift information available, with values spanning the range $2.4\le z_\mathrm{phot}\le11.4$ \citep{2024arXiv240403576K}. The mean and median values of the photometric redshifts is $z_\mathrm{phot}^\mathrm{mean}=7.0$ and $z_\mathrm{phot}^\mathrm{med}=7.2$, respectively. Spectroscopic redshifts are available for $103$ individual LRDs, ranging between $3.1\le z_\mathrm{spec}\le9.1$ \citep{2024arXiv240403576K,2023Natur.616..266L}, with slightly lower mean and median values for this subset, $z_\mathrm{spec}^\mathrm{mean}=6.0$ and $z_\mathrm{spec}^\mathrm{med}=5.7$, respectively. The standard deviations of the redshifts in the subsamples are similar, with $1.7$ and $1.4$ for the photometric and spectroscopic values, respectively. The good photometric redshift estimates were also used for identifying and discarding brown dwarfs from the samples \citep[e.g.][]{2024ApJ...962..177B,2024ApJ...964...39G}. The distributions of spectroscopic and photometric redshift values are illustrated in Fig.~\ref{fig:redshift_dist}.

We cross-matched the positions of LRDs with the latest versions of the quick look catalogues based on the first-epoch (version 3) and second-epoch (version 2) observations of the Very Large Array Sky Survey \citep[VLASS,][]{2020RNAAS...4..175G,2021ApJS..255...30G}. We used two search radii of $1\farcs5$ and $5\arcsec$,  both to look for immediate compact counterparts and to identify more extended emission in the fields of the LRDs.

Additionally, the list of LRD positions were cross-matched with the $1.4$-GHz Faint Images of the Radio Sky at Twenty-centimeters \citep[FIRST,][]{1997ApJ...475..479W,2015ApJ...801...26H} survey data base as well, applying the same search radii. Although the sky coverage, the detection limit ($\sim1$~mJy), and the angular resolution ($5\arcsec$) of FIRST are poorer than those of VLASS, the lower observing frequency could be useful to identify radio counterparts for objects at high redshifts. Despite the poorer angular resolution of FIRST, the search radius was not increased. As it was discussed by \citet{2002AJ....124.2364I}, separations above $2\farcs5$ significantly increase the number of random associations not corresponding to the optical counterpart.


\subsection{Radio counterparts of individual LRDs}
Two radio sources were found within $5\arcsec$ distance to the LRDs in the sample in the FIRST catalogue. We investigated these radio sources individually, and concluded that neither of them are associated with the LRDs in the vicinity (see Appendix~\ref{appendix:radiocounterpart}), and thus all cutouts for the $919$ LRDs were included in the stacking analysis.

\section{Image stacking of LRDs}\label{Results}
We defined empty-field images as cutouts where the signal-to-noise ratio in the $15\arcsec \times 15\arcsec$ area surrounding the position of the LRD was $\mathrm{SNR} \lesssim 6$.

VLASS observed the radio sky  in the $2-4$~GHz frequency range at declinations $\delta\ge-40\degr$. The observations were carried out at three different epochs between $2017$ and $2024$, with the second run of epoch $3$ observations to be completed by the end of 2024. The `Quick Look' radio maps for epochs $1.1-3.1$ are publicly available\footnote{Detailed information on the VLASS observing epochs can be found at \url{https://science.nrao.edu/vlass}} with an angular resolution of $\sim2\farcs5$, and depending on the local root-mean-square (rms) noise levels, the detection threshold ($5\sigma$ rms) can be as low as $\sim300~\mu$Jy. We note that at the time of the analyses,  only the radio maps but not an official component catalogue were available from the third-epoch observational data. Using the \textsc{astroquery} package described in \citet{2019AJ....157...98G}, we searched the VLASS image data base maintained by the Canadian Astronomy Data Centre\footnote{\url{https://www.cadc-ccda.hia-iha.nrc-cnrc.gc.ca/en/}} (CADC). For the three different epochs respectively, we found $1015$, $1015$, and $959$ empty-field `Quick Look’ radio maps centred on $919$, $919$, and $872$ individual LRD positions. Considering the small sample size of LRDs and that the separate images are located on different VLASS tiles, thus they were observed independently at different times, all available maps were included in the stacking.  A total of $770$ maps were acquired from the FIRST survey\footnote{\url{https://sundog.stsci.edu/}}. 

Ideally one should perform the stacking analysis in redshift bins, in order to avoid redshift-dependent biases. Likewise, one should differentiate the sample by source
properties, to minimise selection effects and better characterise the underlying
populations. We performed a stacking in redshift bins, but we did not attempt to
divide LRDs in sub-classes, given the scarcity of observational constraints currently
available for these sources. We divided our sample in four redshift bins and
performed pixel-by-pixel image stacking analyses on the radio maps using both mean and median methods \citep[see e.g.][]{2019MNRAS.490.2542P,2021MNRAS.506.3641G,2024MNRAS.527.3436P}. The redshift-binned images reached $1\sigma$ sensitivities of $\sim5-14~\mu$~Jy~beam$^{-1}$, however, no radio emission was detected (see Appendix~\ref{appendix:bin}). We hence decided to repeat the stacking over the full sample, to improve sensitivity.

\begin{figure*}[ht]
    \centering
    \includegraphics[width=\linewidth]{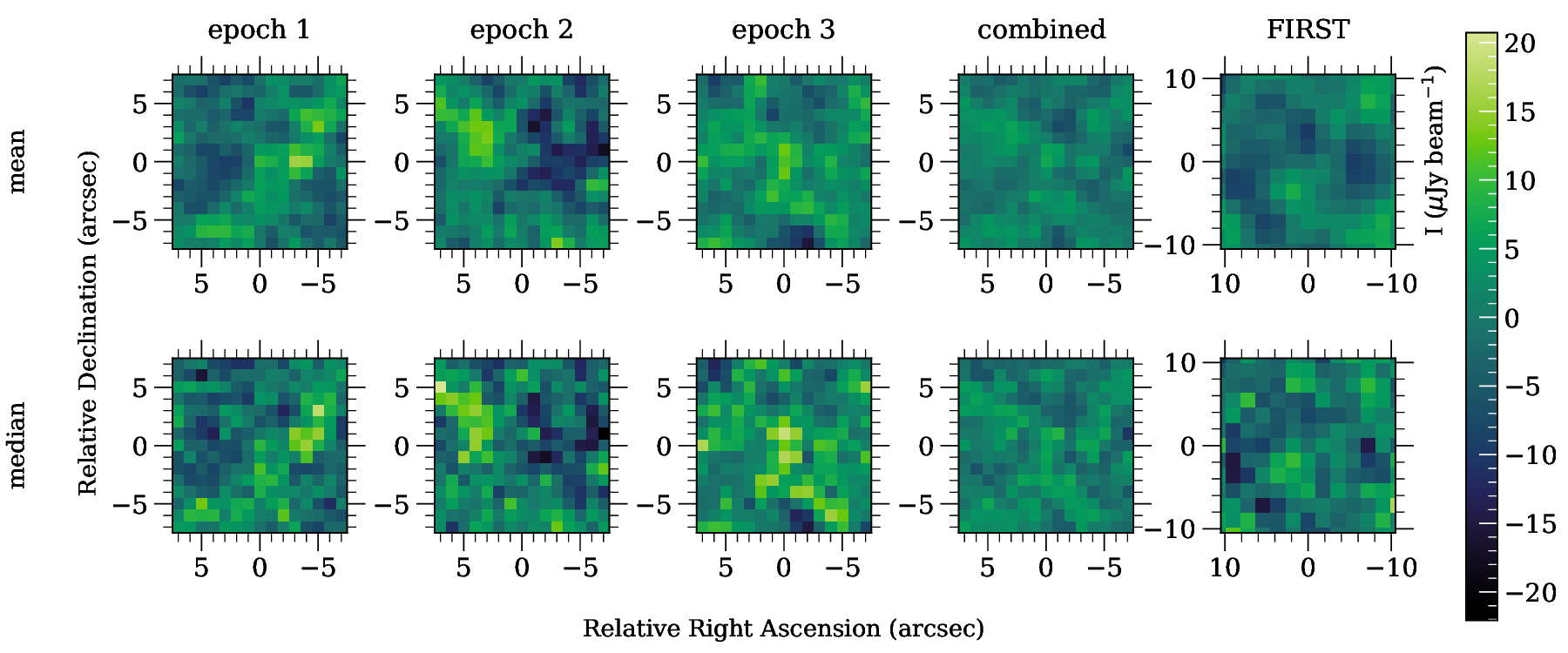}
    \caption{Mean and median stacked empty-field VLASS and FIRST images of the LRDs in the sample. Image properties are listed in Table~\ref{tab:stacked}.}\label{fig:stacked}%
    \end{figure*}

The results of the stacking procedure are illustrated in the panels of Fig.~\ref{fig:stacked}, with the image properties listed in Table~\ref{tab:stacked}. We found no underlying radio emission down to $\sim10~\mu$Jy~beam$^{-1}$-level intensities ($3\sigma$~rms) in either of the stacked images. Upper limits on the radio emission of LRDs were estimated assuming a compact, unresolved structure at arcsec scales. We considered $3\sigma$ rms noise levels for determining flux density values.
The mean and median stacking procedures on the full sample of $2989$ VLASS maps give upper limits on the $3$-GHz flux density as $S_\mathrm{3~GHz}^\mathrm{mean}<8.8~\mu$Jy and $S_\mathrm{3~GHz}^\mathrm{median}<10.8~\mu$Jy, respectively. The mean and median stacked images of the $1.4$-GHz FIRST provide upper limits of $S_\mathrm{1.4~GHz}^\mathrm{mean}<13.3~\mu$Jy and $S_\mathrm{1.4~GHz}^\mathrm{median}<17.7~\mu$Jy. The upper limits on the characteristic monochromatic powers calculated from the median flux density constraints of the stacked VLASS maps give $P_\mathrm{char}^\mathrm{3~GHz} < 1.8\times10^{24}$~W~Hz$^{-1}$, considering the median of all redshift values of the sample, $z^\mathrm{med}_\mathrm{spec+phot}=7.1$. Similarly, the FIRST image upper limits are translated to an upper limit on the characteristic power as $P_\mathrm{char}^\mathrm{1.4~GHz}<2.9 \times10^{24}$~W~Hz$^{-1}$.

    \begin{table}[ht]
        \centering
        \caption{Properties of the mean and median stacked images of LRDs presented in Fig.~\ref{fig:stacked}. Values of the peak intensity ($I_\mathrm{max}$) and the $1\sigma$ rms noise level are given in units of $\mu$Jy~beam$^{-1}$.} 
        \label{tab:stacked}
    \begin{tabular}{ccccccc}
    \hline\hline 
    \multirow{2}{*}{Method} & & \multicolumn{4}{c}{VLASS epoch} & \multirow{2}{*}{FIRST} \\
    	&		&	 1	&	 2	&	 3	&	1+2+3 &  \\	\hline
    	&	maps	&	1015	&	1015	&	959	&	2989	&	770\\	
    	\\											
    \multirow{2}{*}{mean}	&$I_\mathrm{max}$	&	16.0	&	13.8	&	12.7	&	7.0	&	8.5\\	
\multirow{2}{*}{stacking}	&	rms	&	5.2	&	6.5	&	5.1	&	2.9	&	4.4\\
    	&SNR	&	3.1	&	2.1	&	2.5	&	2.4	&	1.9\\	
    	\\												
    \multirow{2}{*}{median}	&$I_\mathrm{max}$	&	18.3	&	20.7	&	19.0	&	8.8	&	16.8\\	
\multirow{2}{*}{stacking}	&	rms	&	6.0	&	7.2	&	6.7	&	3.6	&	5.9\\
    	&SNR	&	3.0	&	2.9	&	2.8	&	2.4	&	2.9	\\	
    
    \hline
    \end{tabular}
    \end{table}

We compared our results with those obtained from the stacking of a sample of $3068$ spectroscopically identifed high-redshift ($4<z<10$) radio-undetected AGNs \citep{2024MNRAS.527.3436P}. The stacking of the control sample of AGNs resulted in similar rms noise levels as for the LRDs in this work, but revealed $\sim30~\mu$Jy flux densities even in single-epoch data sets, with characteristic monochromatic powers in the range ($2\times10^{24}-10^{25}$)~W~Hz$^{-1}$, with a value of $P_\mathrm{char}^\mathrm{AGN}=7\times10^{24}$~W~Hz$^{-1}$ at redshift $z=7.1$. Additionally, it is striking that none of the nearly $1000$ individual LRDs found to date have radio detection. On the contrary, according to our current knowledge, roughly $10-15\%$ of AGNs show radio emission \citep[e.g.][]{2002AJ....124.2364I,2016ApJ...831..168K,2023OJAp....6E..49F}, and in the high-redshift Universe ($z\ge4$) this value was found to be $\sim8\%$ (see the latest version\footnote{The catalogue is available at \url{https://staff.konkoly.hu/perger.krisztina/catalog.html} or \url{https://vizier.cds.unistra.fr/viz-bin/VizieR?-source=J/other/FrASS/4.9}.} of the \citealt{2017FrASS...4....9P} catalogue). This fact and that the upper limit on the monochromatic powers of LRDs is lower than the one found for the comparison high-redshift radio-quiet AGN sample suggest that LRDs host weaker (or no) radio AGN, and/or are a composite population where AGNs are sub-dominant.

Assuming that LRDs are dominated by star formation, the upper limits determined for the radio emission can be translated to star formation rate (SFR) upper limits. Considering that LRDs are indeed high-redshift sources, we applyed the redshift-dependent radio power--SFR correlation of \citet{2017A&A...602A...5N}, resulting in $\mathrm{SFR}<350$~M$_\odot$~yr$^{-1}$ and $\mathrm{SFR}<650$~M$_\odot$~yr$^{-1}$ at the median redshift of the sample ($z=7.1$), depending on the assumed initial mass function, i.e. \citet{2003PASP..115..763C} and \citet{1955ApJ...121..161S}, respectively. We note that applying non-evolving radio power--SFR relations \citep[e.g.][]{2017MNRAS.466.2312D,2019MNRAS.482..560M} would result in values of the same order of magnitude. As galaxies with SFRs in the order of magnitudes of $\sim100$ to $\sim1000$~M$_\odot$~yr$^{-1}$ are not unprecedented  even in the early Universe \citep[e.g.][]{2017ApJ...837..146V,2019ApJ...876...99S,2024A&A...684A..33K}, enhanced star formation as an explanation of the LRD phenomenon cannot be excluded.

\section{Summary and conclusions}
\label{summary}
As to date the underlying physical processes behind the LRD phenomenon are still unknown, we approached the question from a radio viewpoint by collecting data from radio sky surveys at $1.4$ and $3$ GHz, and compared our results with properties of a large sample of high-redshift AGNs. After compiling a list of $919$ LRDs from the literature, we searched for their possible counterparts in the $1.4$-GHz FIRST and $3$-GHz VLASS radio catalogue, and found no associated radio emission, in contrast to the expectations based on the $\sim8\%$ radio-detected fraction of high-redshift AGNs.

A total of $770$ and $2989$ empty-field radio maps centred on the LRD coordinates were acquired from FIRST and VLASS, respectively. Using these image cutouts, we conducted stacking analyses. Considering an unresolved point source as the origin of the non-detected `radio emission', $3\sigma$ flux density upper limits can be estimated for LRDs, and were found to be in the order of $\sim10~\mu$Jy. As stacking of high-redshift $(4 \leq z \lesssim 10)$ radio-quiet AGNs with similar sample size and image noise levels revealed $\sim30~\mu$Jy level flux densities \citep{2019MNRAS.490.2542P,2024MNRAS.527.3436P}, LRDs could represent a population of particularly radio-quiet AGNs. Alternatively, LRDs may be a composite population where AGNs are sub-dominant.

Using the radio non-detection as an upper limit on the flux densities, we estimated the corresponding star formation rates following the radio power-to-SFR relation from \citet{2017A&A...602A...5N}, and found threshold values of $\mathrm{SFR}<350$~M$_\odot$~yr$^{-1}$ and $\mathrm{SFR}<650$~M$_\odot$~yr$^{-1}$.  As SFRs in the order of magnitude of hundreds M$_\odot$~yr$^{-1}$  were oftentimes found in high-redshift galaxies and AGN hosts, LRDs as embodiment of high-redshift galaxies with enhanced star formation are indeed possible. 

The study presented here is the first attempt to characterise the general radio properties of LRDs as a new class of objects at high redshifts. The stacked empty-field radio images suggest significantly weaker radio emission in general, in comparison with a similar high-redshift radio-quiet AGN sample. However, the number of objects that can be stacked at present is only sufficient for deriving an upper limit of the characteristic radio emission. Similar studies in the future, based on much larger LRD samples, would be able to place tighter constraints and perhaps even detect weak stacked radio emission.
Radio image stacking of substantially increased LRD samples may also allow us determining the radio properties of this enigmatic class of objects, as a function of redshift and other important physical parameters.

\begin{acknowledgements}
We thank the referee, Dr. Isabella Prandoni for her constructive suggestions to improve our manuscript.
The National Radio Astronomy Observatory is a facility of the National Science Foundation operated under cooperative agreement by Associated Universities, Inc. CIRADA is funded by a grant from the Canada Foundation for Innovation 2017 Innovation Fund (Project 35999), as well as by the Provinces of Ontario, British Columbia, Alberta, Manitoba and Quebec. We thank the Hungarian National Research, Development and Innovation Office (NKFIH, grants OTKA K134213 and PD146947) for support. This work was also supported by HUN-REN and the NKFIH excellence grant TKP2021-NKTA-64.
\end{acknowledgements}

\bibliographystyle{aa.bst}
\bibliography{lrd_radio.bib}

\begin{appendix}
\section{Radio counterparts of individual LRDs}\label{appendix:radiocounterpart}
\begin{figure}[ht]
    \centering
    \includegraphics[width=\linewidth]{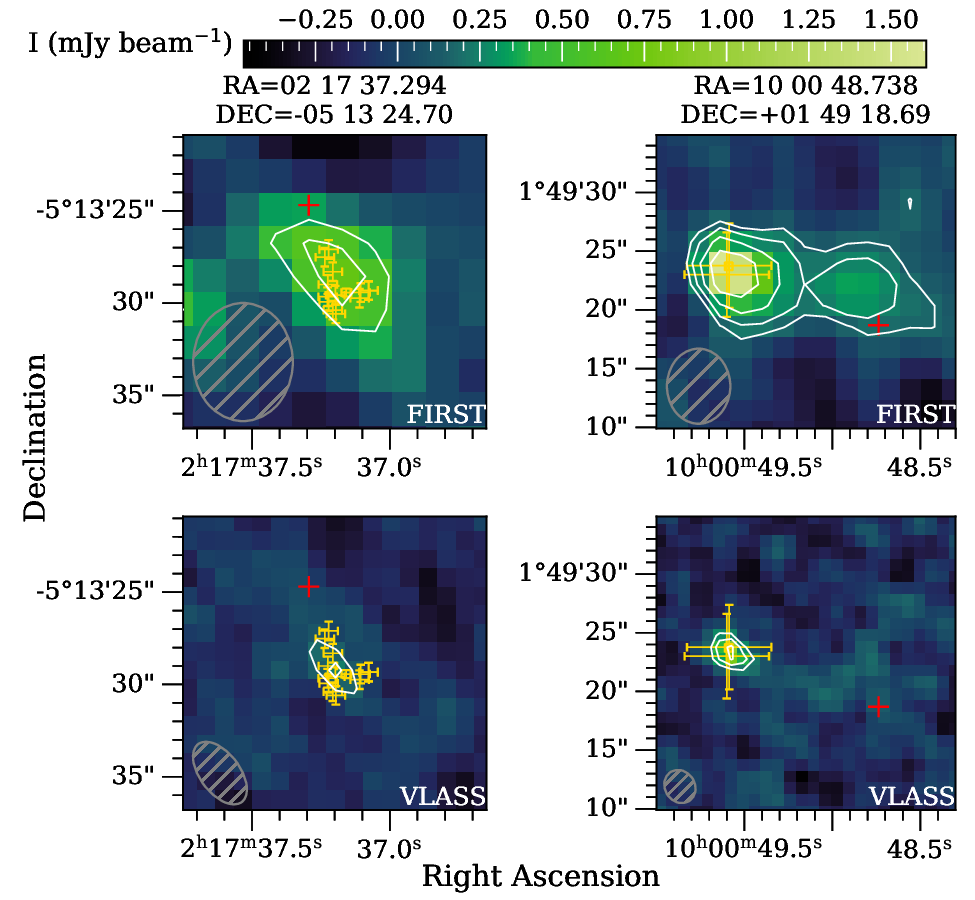}
    \caption{FIRST and VLASS image cutouts around two LRDs, J0217$-$0513 (id. 52581, left) and J1000+0149 (id. 381494, right). The coordinates corresponding to the positions of the LRDs (red crosses) are shown on the top of each column. Yellow markers denote the most probable optical counterparts of the radio emission, the gravitational lensing system SL2S~J02176$-$0513 \citep{2009A&A...501..475T} for the first, and a radio galaxy \citep{2010ApJ...713..484O} or a galaxy cluster \citep{2010ApJ...714..218G} for the second radio source. The first radio contours (white) are drawn at $\pm3\sigma$ rms noise levels ($\sim0.4$~mJy~beam$^{-1}$), and the positive levels increase by a factor of $\sqrt{2}$. The restoring elliptical Gaussian beam sizes (half-power beam widths) are shown in the bottom left corners ($5\farcs4\times6\farcs4$ at major axis position angle $\mathrm{PA}=0\degr$ for the FIRST maps, and $2\farcs2\times3\farcs9$ at $\mathrm{PA}=37\degr$ and $2\farcs6\times2\farcs9$ at $\mathrm{PA}=33\degr$ for the left and right VLASS images, respectively).} \label{fig:sourceclose}%
\end{figure}

In the FIRST catalogue, only two radio sources were found within $5\arcsec$ distance to the LRDs in the sample, and one of those was also identified in the first-epoch VLASS radio component catalogue, in the vicinity of the LRDs J0217$-$0513 \citep[id. 52581,][]{2024ApJ...968...38K}, and J1000+0149 \citep[id. 381494,][]{2024arXiv240610341A}. The latter remained `undetected' in the VLASS catalogue. This appears to be the fainter component of a double radio source, with the brighter component being at $11\farcs7$ distance. No counterparts were identified in the second-epoch VLASS catalogue, although both radio sources are visible in all three epochs of the `Quick Look' images. This can be attributed to the low signal-to-noise ratio of $\mathrm{SNR} \lesssim 5$, which is just below the VLASS-defined detection limit.  These radio sources are shown in Fig.~\ref{fig:sourceclose}. 

The finer angular resolution of VLASS, the astrometric precision of \textit{JWST}, and the presence of known optical sources coinciding with the radio coordinates strongly suggest that the LRDs are unlikely to be connected to the radio emission. The radio source closest to J0217$-$0513 is surrounded with the components of the gravitational lensing system SL2S~J02176$-$0513, with the lens galaxy ($z=0.656$) coinciding with the radio peak, and the arc-shaped images of the lensed galaxy ($z=1.847$) `enveloping' the eastern side of the radio feature \citep{2009A&A...501..475T}. The  brighter component of the double radio source in the field of J100+0149 is consistent with the position of both a galaxy \citep{2010ApJ...713..484O} and a galaxy cluster \citep{2010ApJ...714..218G} at $z=0.53$. The fact that the LRD is unrelated to the radio emission from the foreground sources is also supported by the radio morphology seen in the more sensitive $1.3$-GHz radio map from recent MeerKAT observations \citep{2022MNRAS.509.2150H}.

Based on the above, we can conclude that none of the $919$ LRDs in our sample can be securely associated with any radio source in the VLASS and FIRST surveys.
 
\section{Stacking redshift-binned subsamples}\label{appendix:bin}

 \begin{figure*}[hpt]
    \centering
    \includegraphics[width=0.75\linewidth]{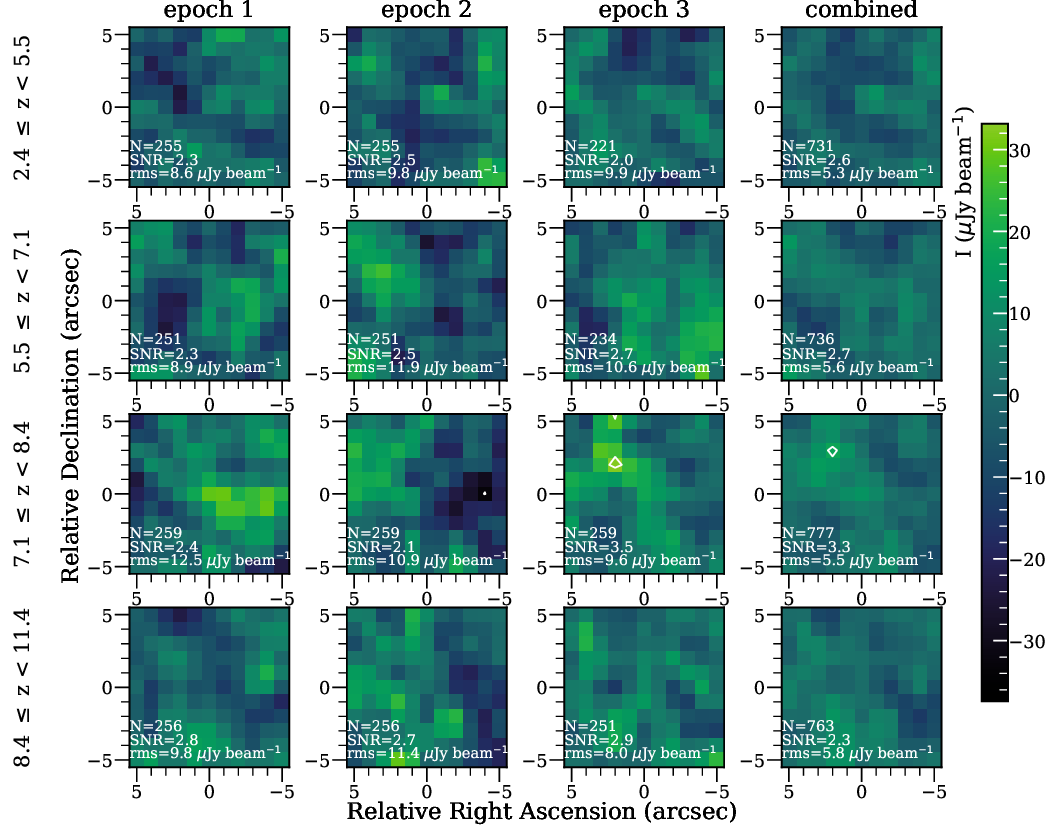}
    \caption{Mean stacked VLASS images of the LRDs in the sample, divided into four redshift bins. Contour lines denote the $\pm3\sigma$ rms noise levels.}\label{fig:stackedbinmean}%
\end{figure*}

\begin{figure*}[ht]
    \centering
    \includegraphics[width=0.75\linewidth]{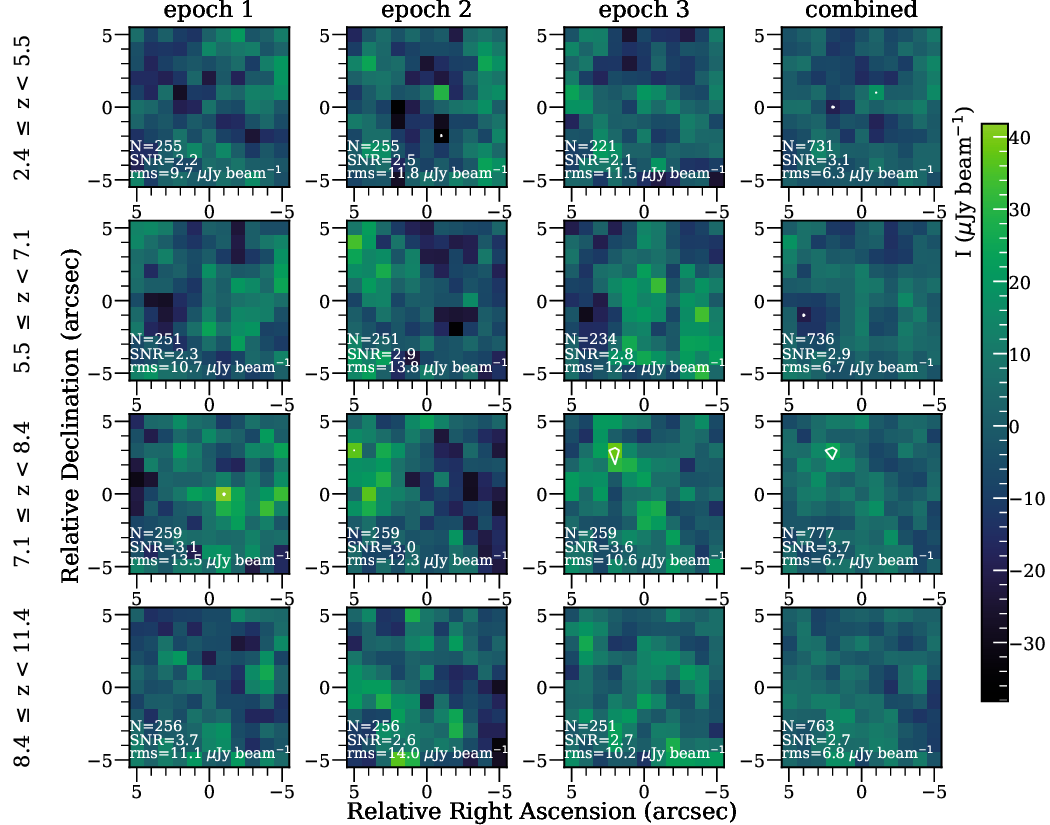}
    \caption{Median stacked VLASS images of the LRDs in the sample, divided into four redshift bins. Contour lines denote the $\pm3\sigma$ rms noise levels.}\label{fig:stackedbinmedian}%
\end{figure*}

As of today, the exact physical properties and any statistically significant differences in the LRD population are still under examination. However, it is likely that LRDs represent various kinds of different objects, and thus show different properties depending on the redshift. Stacking analyses were carried out with dividing the LRDs into four redshift bins with approximately equal number of sources in each. As the decrease of the image noise follows the $1/\sqrt{N}$ relation, where $N$ is the number of objects in the stacked sample, defining subsamples would limit the sensitivity of the stacked images. Indeed, we found higher rms levels ($\sim5-14~\mu$Jy~beam$^{-1}$) with similarly low  SNR ($\sim2-3$) values in both the mean (Fig~\ref{fig:stackedbinmean}) and median (Fig~\ref{fig:stackedbinmedian}) stacked VLASS maps in of the redshift-binned sample. 
Based on the results on both the mean and median-stacked binned-data, no further conclusions can be drawn. Thus, at the moment we can only discuss the stacking of the sample as a whole in detail.

\end{appendix}

\end{document}